\documentclass{sigcomm-alternate}

\usepackage{algorithm}
\usepackage{algorithmic}
\usepackage{graphicx}
\usepackage{subcaption}
\usepackage[hyphens]{url}
\usepackage{booktabs}
\usepackage{multirow}
\usepackage{color}

\usepackage{enumitem}

\newcommand{\ignore}[1]{}

\newcommand{\eg}{{\it e.g.,}~}
\newcommand{\ie}{{\it i.e.,}~}
\newcommand{\cf}{{\it c.f.,}~}
 
\newcommand{\figref}[1]{Figure~\ref{#1}}

\def\authnotes{1}
\newcommand{\authnote}[2]{\ifnum\authnotes=1\begin{quote}\textcolor{blue}{\textbf{#1 says:} #2}\end{quote}\fi}

\begin{document}

\title{Privacy Failures in Encrypted Messaging Services:\\Apple iMessage and Beyond}
\numberofauthors{2}

\author{
	Scott E. Coull\\
	\affaddr{RedJack, LLC.}\\
	\affaddr{Silver Spring, MD}\\
	\email{\normalsize scott.coull@redjack.com}
	\and
	Kevin P. Dyer\\
	\affaddr{Portland State University}\\
	\affaddr{Portland, OR}\\
	\email{\normalsize kdyer@cs.pdx.edu}
}

\maketitle

\begin{abstract}
Instant messaging services are quickly becoming the most dominant 
form of communication among consumers around the world.  Apple iMessage, for example, handles over 2 billion 
message each day, while WhatsApp claims 16 billion messages from 400 million international users.  To protect user 
privacy, these services typically implement end-to-end and transport layer encryption, which are meant to 
make eavesdropping infeasible even for the service providers themselves.  In this paper, however, we show that it is 
possible for an eavesdropper to learn information about user actions, the language of messages, and even the length of those messages with greater than 96\% accuracy 
despite the use of state-of-the-art encryption technologies simply by observing the sizes of encrypted packet.  
While our evaluation focuses on Apple iMessage, the attacks are completely generic and 
we show how they can be applied to many popular messaging services, including WhatsApp, Viber, and Telegram.
\end{abstract}

\section{Introduction}
\label{sec:intro}
Over the course of the past decade, instant messaging services have gone from a niche application used on desktop computers 
to the most prevalent form of communication in the world, due in large part to the growth of Internet-enabled phones and tablets.  Messaging 
services, like Apple iMessage, Telegram, WhatsApp, and Viber, handle \emph{tens of billions} of messages each day from an international user 
base of over one billion people \cite{Lovejoy:Apple_Growth,Olson:WhatsApp}.  Given the volume of messages traversing these services and 
ongoing concerns over widespread eavesdropping of Internet communications, it is not surprising that privacy has been an important topic for 
both the users and service providers.  To protect user privacy, these messaging services offer transport layer encryption technologies to protect 
messages in transit, and some services, like iMessage and Telegram, offer end-to-end encryption to ensure that not even the 
providers themselves can eavesdrop on the messages \cite{Apple:iOS_Security,Greenberg:Apple_PRISM}.
As previous experience with Voice-over-IP (\eg \cite{White:VoIP,Wright:VoIP_Spot}) and HTTP tunnels (\eg \cite{Dyer:Peekaboo, Panchenko2011}) 
has shown us, however, the use of state-of-the-art encryption technologies is no guarantee of 
privacy for the underlying message content.

In this paper, we analyze the network traffic of popular encrypted messaging services to (1) understand the breadth 
and depth of their information leakage, (2) determine if attacks are generalizable across services, and (3) calculate the potential costs of protecting against 
this leakage.  Specifically, we focus our analysis on the Apple iMessage service and show that it is possible to reveal information about the device operating system, 
fine-grained user actions, the language of the messages, and even the approximate message length with accuracy exceeding 96\%, as shown in the  
summary provided in Table \ref{tbl:summary}.  In addition, we demonstrate that these attacks are applicable to many other popular messaging services, 
such as WhatsApp, Viber, and Telegram, because they target deterministic relationships between user actions and the resultant encrypted packets that exist 
regardless of the underlying encryption methods or protocols used.  Our analysis of countermeasures shows that the attacks can be completely 
mitigated by adding random padding to the messages, but at a cost of over 300\% overhead, which translates to at least a \emph{terabyte} of 
extra data \emph{per day} for the service providers.
Overall, these attacks could impact over a billion users across the globe and the high level of accuracy that we demonstrate 
in our experiments means that they represent realistic threats to privacy, particularly given recent revelations about widespread metadata collection by government 
agencies \cite{Cohn:Metadata}.

\begin{table}[t!]
\center
\begin{tabular}{ l l r }
  \textbf{Attack} & \textbf{Method} & \textbf{Accuracy} \\
  \hline
  Operating System & Na\"ive Bayes & 100\% \\
  User Action & Lookup Table & 96\% \\
  Language &  Na\"ive Bayes& 98\% \\
  Message Length & Linear Regression & 6.27 chars.\\
\end{tabular}
\caption{Summary of attack results for Apple iMessage.}
\label{tbl:summary}
\end{table}

\ignore{
\begin{figure*}[t!]
	\footnotesize
	\centering
	\begin{subfigure}{\columnwidth}
		\footnotesize
		\centering
		\includegraphics[width=0.99\textwidth]{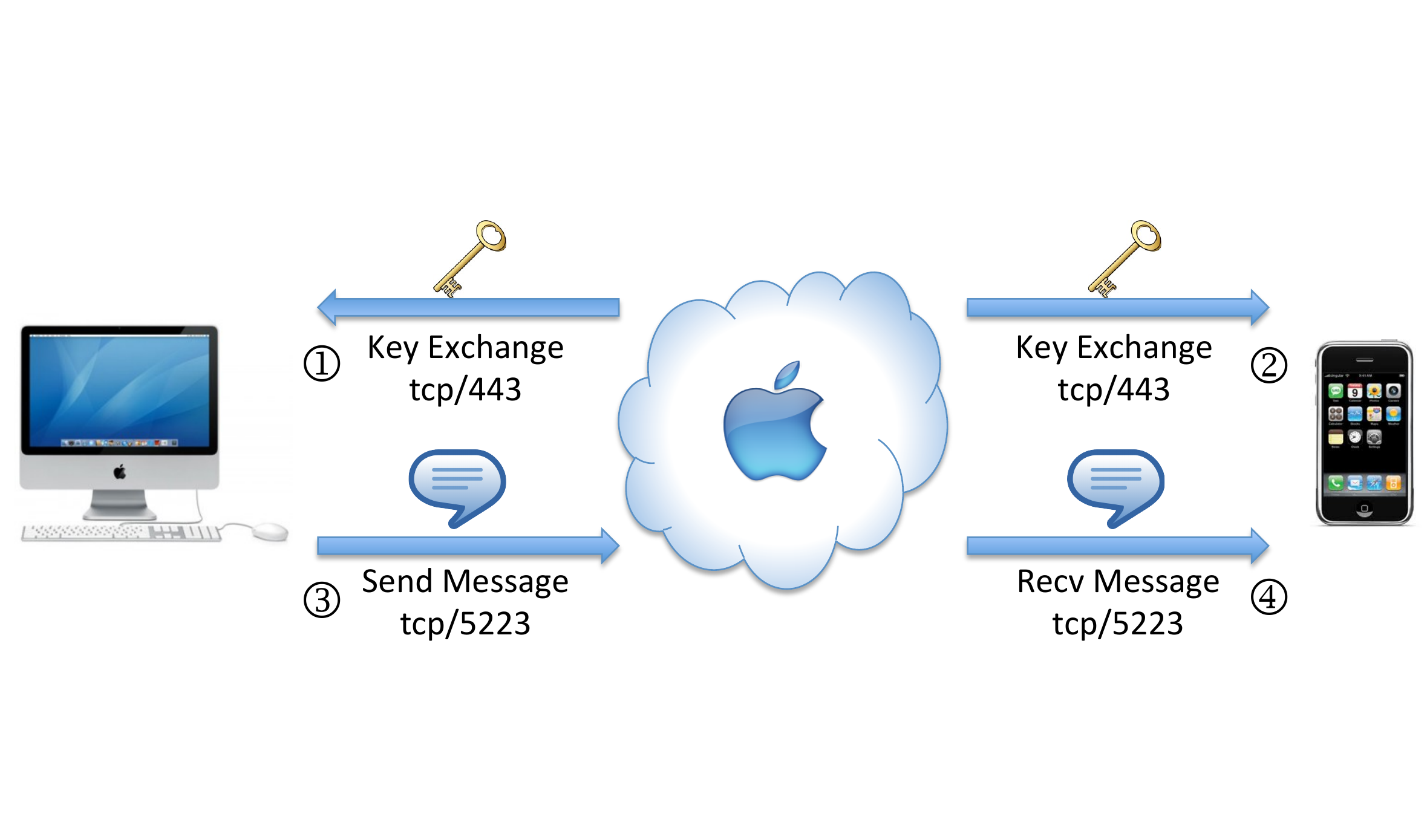}
		\caption{Sending a text message}
		\label{fig:send_message}
	\end{subfigure}
	\hfill
	\begin{subfigure}{\columnwidth}
		\footnotesize
		\centering
		\includegraphics[width=0.99\textwidth]{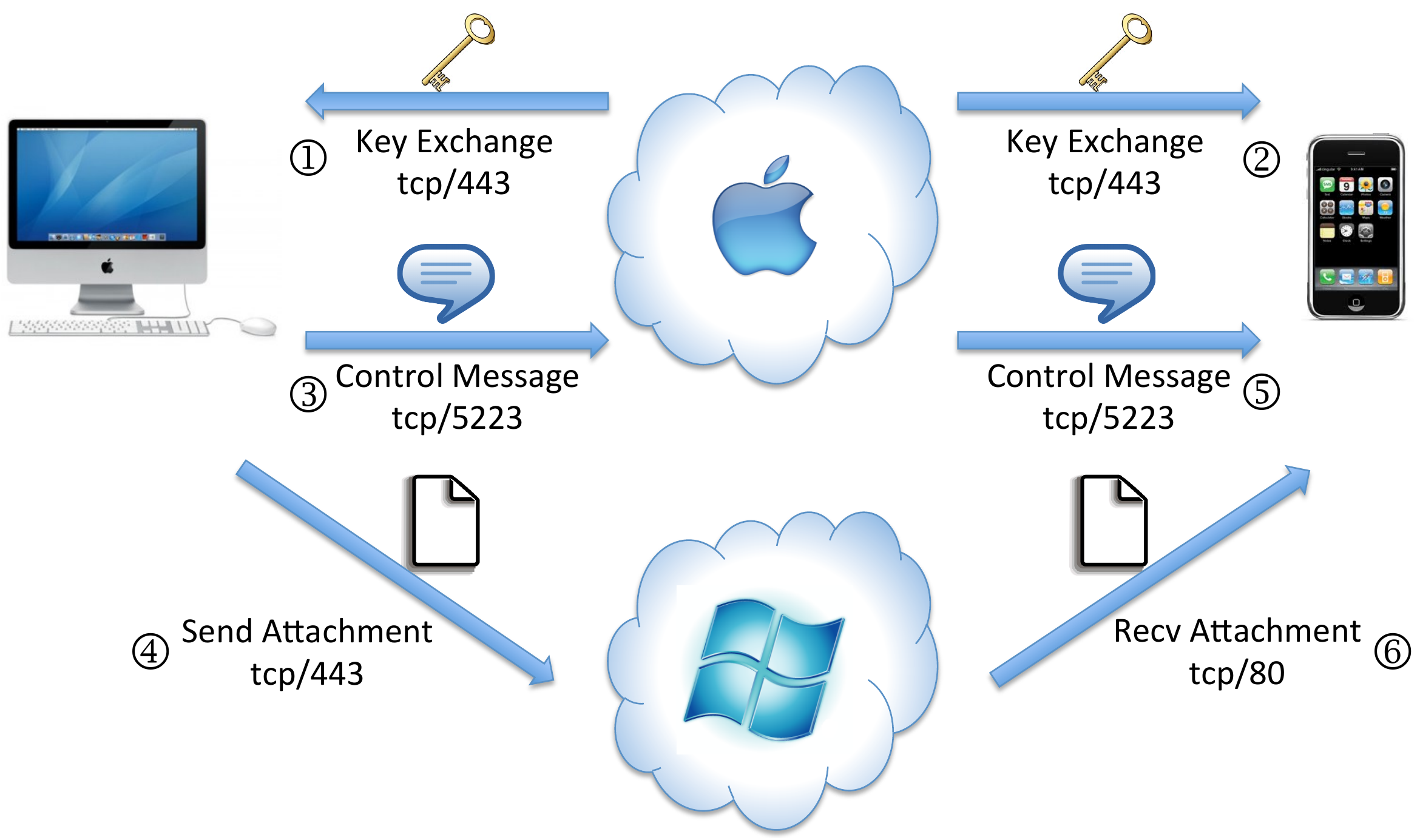}
		\caption{Sending an attachment}
		\label{fig:send_attach}
	\end{subfigure}
	\caption{ Protocol operation for (a) sending a text message and (b) an attachment.  Arrows indicate primary direction of data flow.}
	\label{fig:proto_overview}
\end{figure*}
}

\section{Background}
\label{sec:background}
Before we begin our analysis, we first provide an overview of the iMessage service, and discuss prior work 
in the analysis of encrypted network traffic.  Interested readers should refer to documentation from projects focused 
on reverse engineering specific portions of the iMessage service \cite{PushProxy, Goodin:iMessage_Crypto, Green:iMessage_Crypto}, 
or the official Apple iOS security white paper \cite{Apple:iOS_Security}.

\subsection{iMessage Overview}
iMessage uses the Apple Push Notification Service (APNS) to deliver text messages and attachments to users.  
When the device is first 
registered with Apple, a client certificate is created and stored on the device.  Every time the device is connected 
to the Internet, a persistent APNS connection is made to Apple over TCP port 5223.  The connection appears to be 
a standard TLS tunnel protecting the APNS messages.  From here, the persistent APNS connection is used to send and receive 
both control messages and user content for the iMessage service.  If the user has not recently interacted with the sender or recipient of 
a message, then the client initiates a new TLS connection with Apple on TCP port 443 and receives key information 
for the opposite party.  Unlike earlier TLS connections, this one is authenticated using the client certificate generated 
during the registration process.
Once the keys are established, there are five user actions that are observable through the APNS and TLS connections 
made by the iMessage service.  These actions include:  (1) start typing, (2) stop typing, (3) send text, (4) send attachment, 
and (5) read receipt.  All of the user 
actions mentioned follow the protocol flow shown in \figref{fig:send_message}, except for sending an attachment.  
The protocol flow for attachments is quite similar except that the attachment itself is stored 
in the Microsoft Azure cloud storage system before it is retrieved, rather than being sent directly through Apple.

\begin{figure}[t]
	\scriptsize
	\center
	\includegraphics[width=0.95\columnwidth]{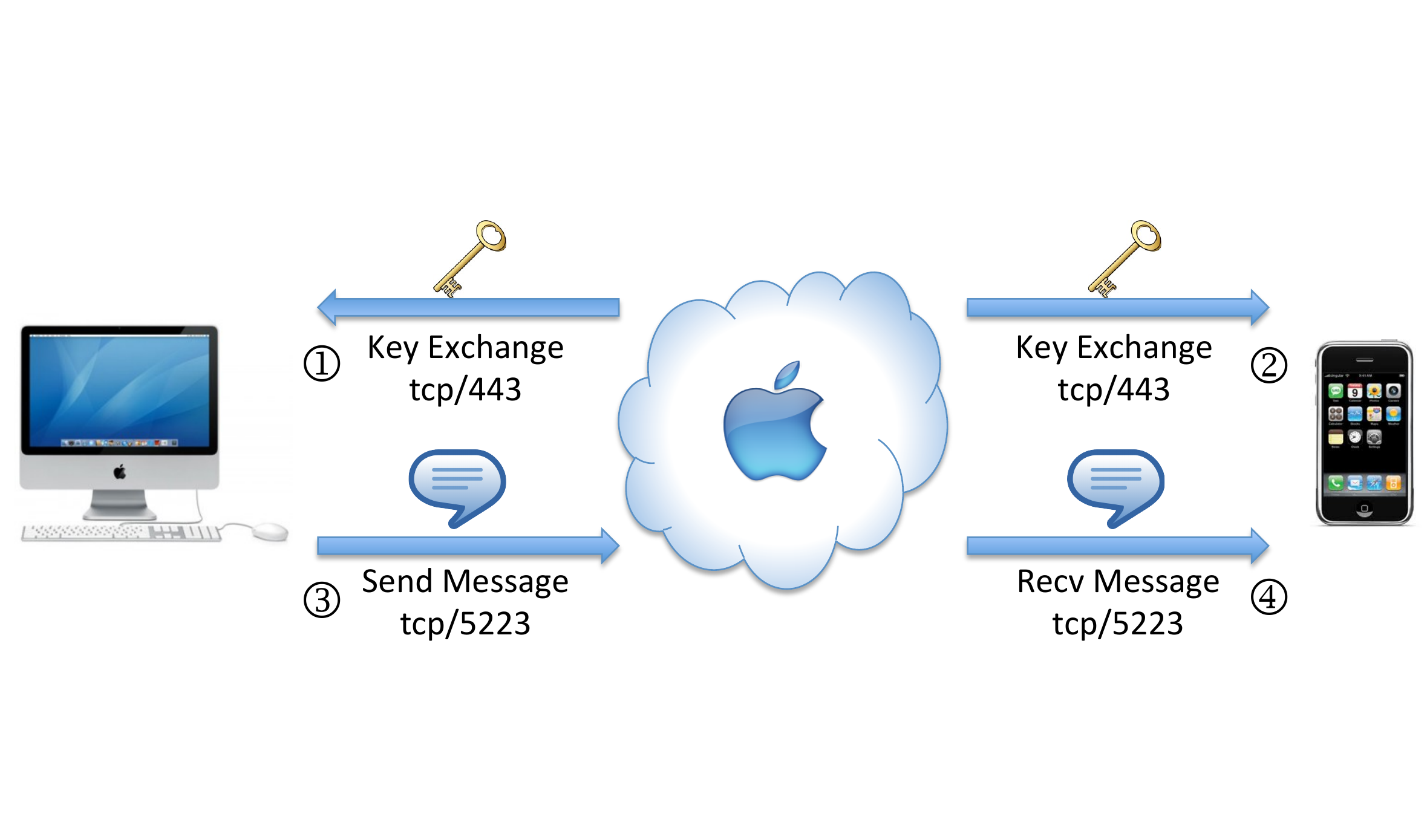}
	\caption{High-level operation of iMessage.}
	\label{fig:send_message}
\end{figure}

Over the course of our analysis, we observed some interesting deviations from this standard 
protocol.  For instance, when TCP port 5223 is blocked, the APNS message stream shifts to using TCP port 443.  
Similarly, cellular-enabled iOS devices use port 5223 while connected to the cellular network, but 
switch to port 443 when WiFi is used.  Moreover, if the iOS device began its connection using the cellular network, that connection 
will remain active even if the device is subsequently connected to a wireless access point.  It is important to note 
that payload sizes and general APNS protocol behaviors remain exactly the same regardless of if port 5223 or 443 are used, and 
therefore any attacks on the standard APNS scenarios are equally applicable in both cases.  

\subsection{Related Work}
To date, there have been two primary efforts in understanding the operation of the iMessage service and the APNS protocol.  Frister and 
Kreichgauer have developed the open source Push Proxy project \cite{PushProxy}, which allows users to decode APNS messages into a readable 
format by redirecting those messages through a man-in-the-middle proxy.  
In another recent effort, Matthew Green \cite{Green:iMessage_Crypto} and Ashkan Soltani \cite{Goodin:iMessage_Crypto} showed that, 
while iMessage data is protected by end-to-end encryption, the keys used to perform that 
encryption are mediated by an Apple-run directory service that could potentially be used by an attacker (or Apple themselves) 
to install their own keys for eavesdropping purposes.  
More broadly, the techniques presented in this paper follow from a long line of attacks that use only the timing and size of encrypted network traffic 
to reveal surprising amounts of information.  In the past, traffic analysis methods have been applied in identifying web pages \cite{Dyer:Peekaboo,
Herrmann2009, Liberatore:Web, Panchenko2011,Sun:EncWeb}, and reconstructing 
spoken phrases in VoIP \cite{White:VoIP,Wright:VoIP_Spot}.

To the best of our knowledge, this is the first paper to examine the privacy of encrypted instant messaging services, particularly those used by mobile devices.  
We distinguish ourselves from earlier work in both the broad impact and realistic nature of our attacks.  Specifically, we demonstrate highly-accurate attacks that 
could affect nearly a billion users across a wide variety of messaging services, whereas previous work in other areas of encrypted traffic analysis have relatively small 
impact due to limited user base or poor accuracy.  
When compared to earlier work in analyzing iMessage, we focus on an eavesdropping scenario that requires no cooperation from service providers  
and has been demonstrated to exist in practice \cite{Cohn:Metadata}.

\section{Analyzing Information Leakage}

\label{sec:analyze}
In this section, we investigate information leakage about devices, users, and messages by analyzing 
the relationship between packet sizes within the persistent APNS connection used by iMessage and 
user actions.  For each of these categories of leakage, we first provide a general analysis of the data 
to discover trends or distinguishing features, then evaluate classification strategies capable 
of exploiting those features.

\subsection{Data and Methodology}
\label{sec:data_method}
To evaluate our classifiers, we collected data for each of the five observable user actions (start, stop, 
text, attachment, read) by using scripting techniques that drove the actual iMessage user interfaces 
on OSX and iOS devices.  Specifically, we used Applescript to natively type text, paste images, and send/read  
messages on a Macbook Pro running OSX 10.9.1, and a combination of VNC remote 
control software and Applescript to control the same actions on a jailbroken iPhone 4 (iOS 6.1.4).  For each 
user action, we collected 250 packet capture examples on both devices and in both directions of communications 
(\ie to/from Apple) 
for a total of 5,000 samples.   
In addition, we also collected small samples of data using devices running iOS 5, iOS 7, and OSX Mountain Lion to verify the observed 
trends.

\ignore{
\begin{figure}
	\footnotesize
	\centering
	\includegraphics[width=0.98\columnwidth]{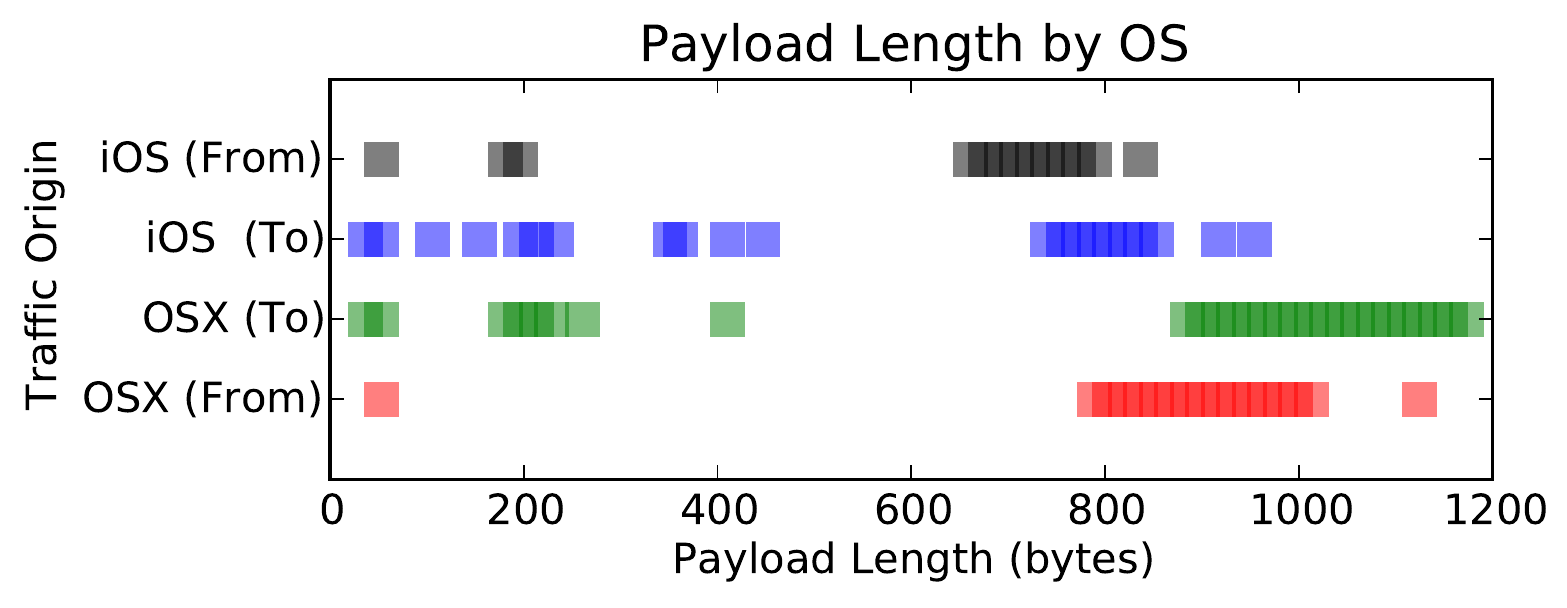}
	\caption{Payload lengths by OS and direction.}
	\label{fig:line_os}
\end{figure}
}

\begin{figure}
	\footnotesize
	\centering
	\includegraphics[scale=0.5]{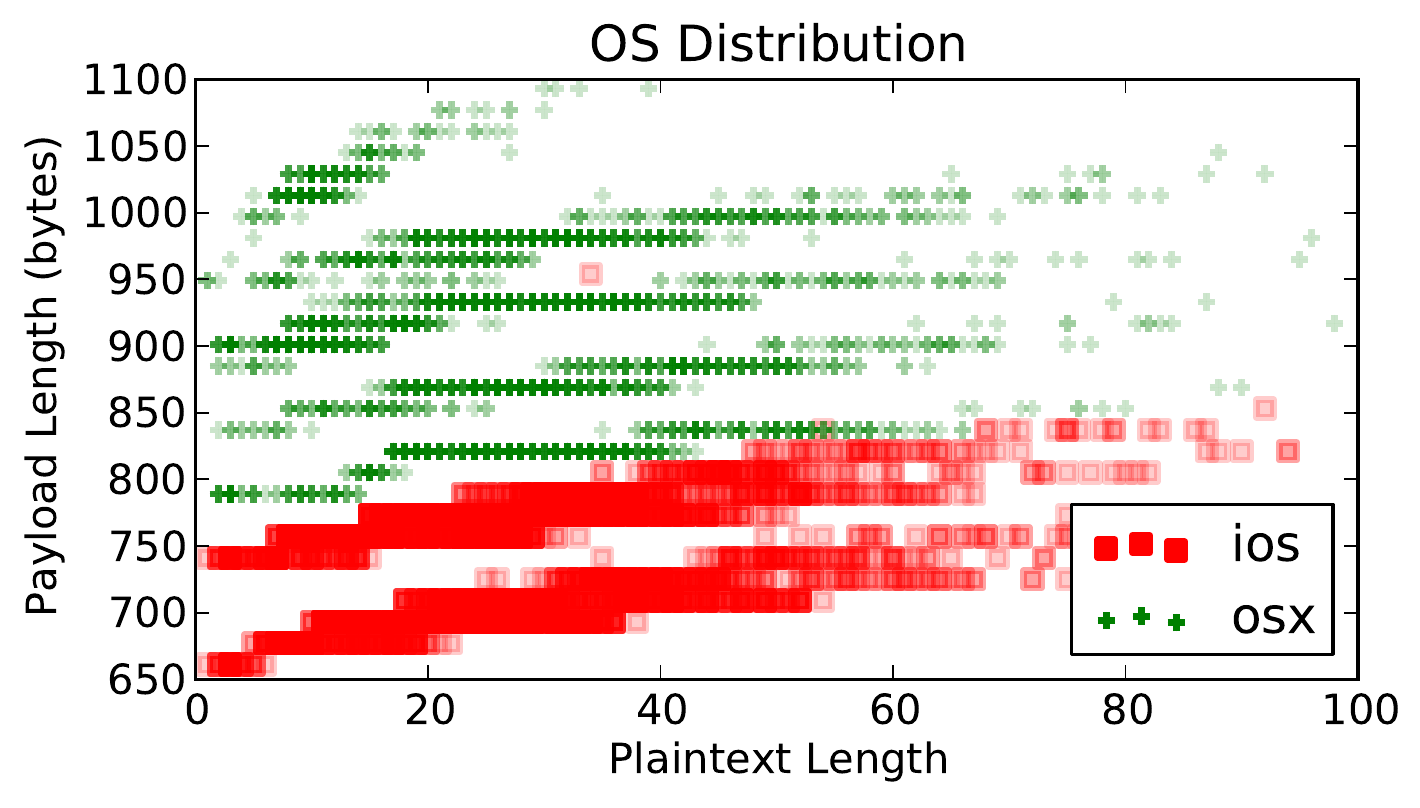}
	\caption{Scatter plot of plaintext message lengths versus ciphertext lengths for packets containing user content.}
	\label{fig:scatter_os}
\end{figure}

\begin{figure*}[t!]
\footnotesize
\centering
\includegraphics[scale=0.5]{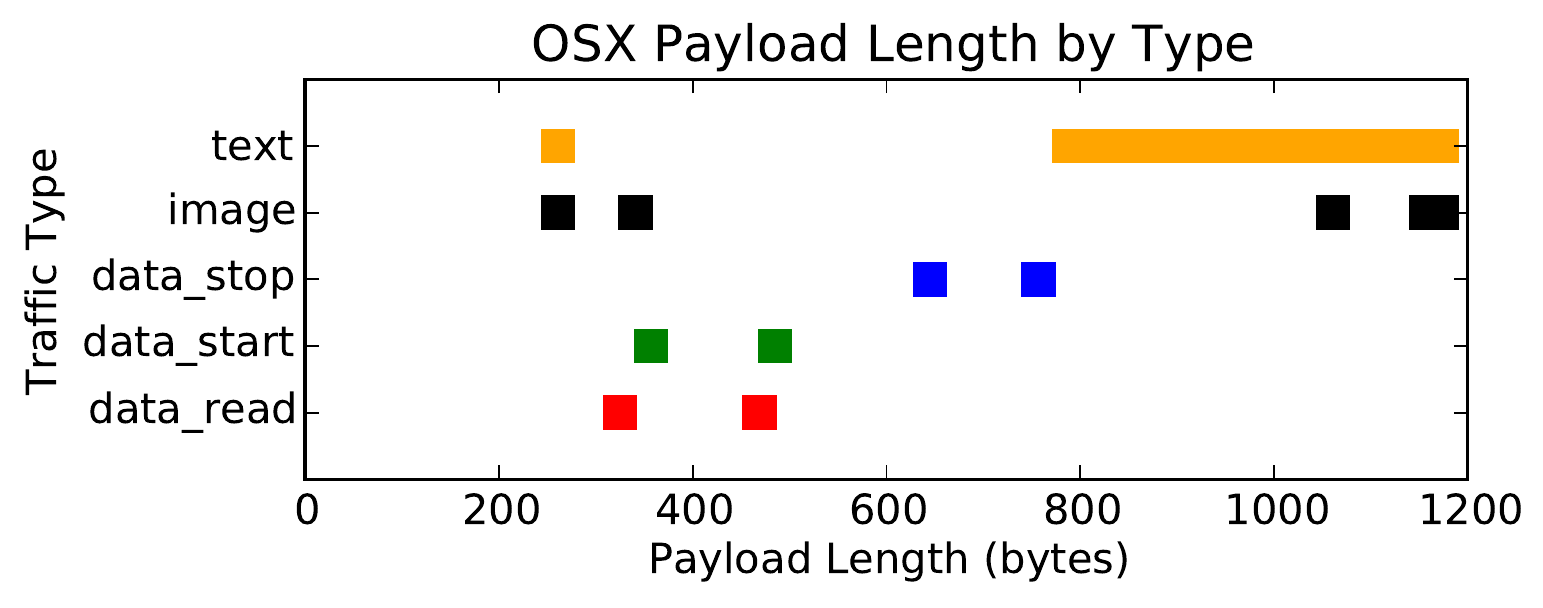}
\includegraphics[scale=0.5]{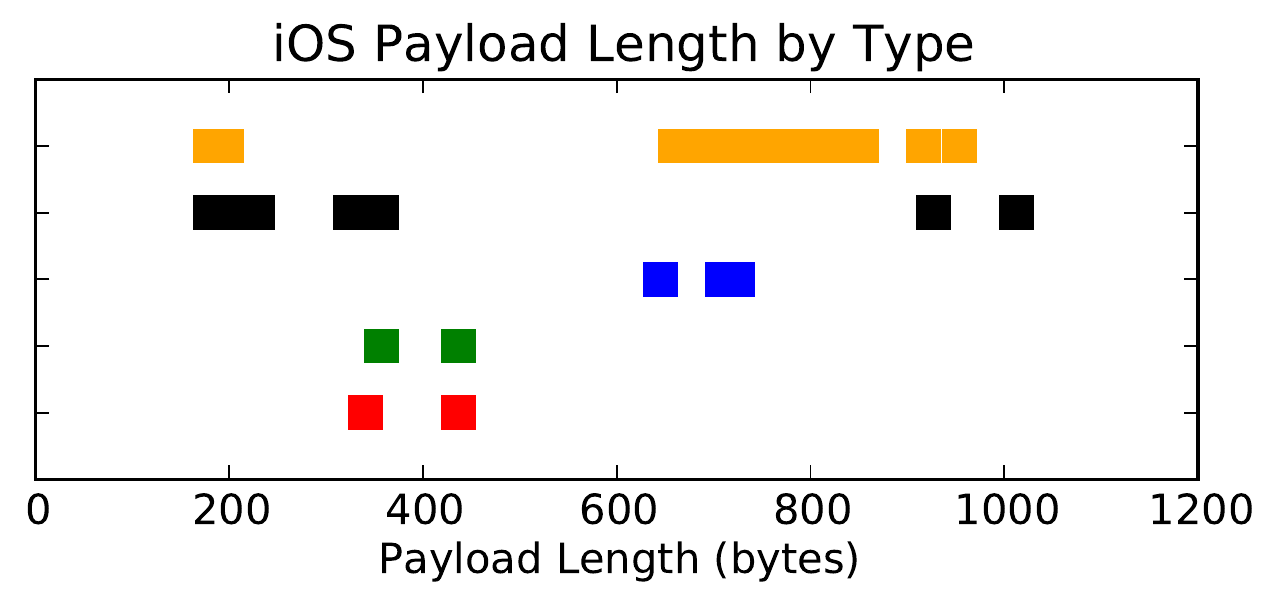}
\caption{Distribution of payload lengths for each message type separated by operating system without control packets.}
\label{fig:line_types}
\end{figure*}

The underlying text data is drawn from a  set of over one million sentences and short 
phrases in a variety of languages from the Tatoeba parallel translation corpus \cite{Tatoeba}.  Languages used 
in our evaluation include Chinese, English, French, German, Russian, and Spanish.  For attachment data, 
we randomly generated PNG images of exponentially increasing size (64 x 64, 128 x 128, 256 x 256).  Throughout the 
remainder of the paper, we simply refer to attachments as ``image" messages.  
Although the Tatoeba dataset does not contain typical text message shorthand, it is generated 
through a community of non-expert users (\ie crowd-sourced) and so actually contains several informal phrases 
that are not found in a typical language translation corpus.

Each experiment in this section used 10-fold cross validation testing, where the data for each 
instance in the test was constructed by sampling TCP payload lengths and packet directions (\ie to/from Apple) 
from the relevant subset of the packet capture files.  The only preprocessing that was performed on the data was 
to remove duplicate packets that occur as a result of TCP retransmissions and those packets without TCP payloads.  
Performance of our classifiers is report with respect to overall accuracy, which is 
calculated as the sum of the true positives and true negatives over the total number of samples evaluated.  
Where appropriate, we also use confusion matrices that show how each of the test instances 
was classified and use absolute error to measure the predictive error in our regression analysis.  
%A full description of the na\"ive Bayes and linear regression classifiers that we use can be found in any standard machine learning textbook \cite{Mitchell1997}.

\subsection{Operating System}
\label{sec:fingerprint}
Our first experiment examines the difference in the observable packet sizes for the iOS and OSX operating systems.  
The scatterplot of iMessage packet sizes in \figref{fig:scatter_os}  
shows how iOS appears to more efficiently compress the plaintext, while OSX occupies a much larger space.  These two classes 
of data are clearly separable, but the figure also shows five unique bands of plaintext/ciphertext relationship, which hints at leakage of 
finer-grained information about the individual messages (which we examine in Section \ref{sec:lang}).
Additionally, when we break down the distributions based on their direction (to/from Apple), we see that there is a deterministic 
relationship between the two.  That is, as messages pass through Apple, 112 bytes of data are removed from OSX messages and 
64 bytes are removed from iOS messages.  Aside from the ability to fingerprint the OS version, the deterministic nature of 
these changes indicates that it is also possible to correlate and trace communications as it passes through Apple on the way to its destination.

To identify the OS of observed devices, we use a binomial na\"ive Bayes classifier from the 
Weka machine learning library \cite{Weka} with one class for each of the four possible OS, direction combinations.  
The classifier operates on a binary feature vector of packet length, direction pairs, where 
the value for a given dimension is set to ``true" if that pair was observed and ``false" otherwise.  
To determine the number of packet observations necessary for accurate classification, we run 10-fold cross-validation 
experiments where the 1,024 instances used for each experiment are created with $N=1,2,\ldots,50$ packets sampled from the appropriate subset of the dataset 
for each OS, observation point class.  
The results indicate that we are able to accurately classify the OS with 100\% accuracy after observing only five packets regardless of the 
operating system.  A cursory analysis of iOS 5 and 7 indicates that they also produce messages with lengths that are unique 
from both the OSX and iOS 6.1.4 device, which indicates that this type of device fingerprinting could be refined to reveal specific version information
when the size of the APNS messages changes between OS versions.

%Practically speaking, we should be concerned about two uses of this type of information.  First, since the message sizes change in a deterministic fashion as they pass through 
%Apple, it may be possible to correlate communications between two end users simply by matching the timing and sizes of the packets as they are observed entering and 
%leaving the Apple servers.  Second, since different operating systems and versions are observable through the distribution of packet sizes, it is possible for an eavesdropper 
%to target and exploit vulnerable devices of her choosing.

\subsection{User Actions}
\label{sec:types}
Recall from our earlier 
discussion that there are five high-level user actions that we can observe:  start, stop, text, 
attachment (image), and read.  \figref{fig:line_types} 
shows the distribution of payload lengths for each of these actions separated by 
the OS of the sending device after removing control packets (\ie packet sizes that occur within multiple classes).  
Most classes have two distinctive packet lengths -- one for when the message is sent to Apple and one when it is 
received from Apple.
The only classes that overlap substantially are the read receipt and start messages in the iOS data going to Apple.

The stability and deterministic nature of the payload lengths in most classes makes the use of probabilistic classifiers unnecessary.   
Instead of using 
heavyweight machine learning methods, we create a hash-based lookup table using each observed length 
in the training data as a key and store the associated class labels.  
In addition to creating 
classes for the five standard message types derived from user actions, we also create a class for the 
payload lengths of identified control packets.  When a new packet arrives, we check the lookup table to retrieve 
the class label(s) for its payload length.  If only one label is found, the packet is labeled as that message type.  
In the case where two class labels are returned, we choose the class where that payload length occurs most 
frequently in the training data.

\begin{table*}
	\scriptsize
	\begin{center}
		\begin{tabular}{ccccccccccccc}
			%\toprule
			\multicolumn{6}{c}{OSX (From)} & & \multicolumn{6}{c}{OSX (To)}\\
			%\cline{1-7} \cline{9-15}
			\cline{1-6} \cline{8-13}
			control & read & start & stop & image & text &  & control & read & start & stop & image & text \\
			\midrule
			\textbf{1.0} & 0.0 & 0.0 & 0.0 & 0.0 & 0.0 &control &\textbf{1.0} & 0.0 & 0.0 & 0.0 & 0.0 & 0.0 \\
			0.0 & \textbf{1.0} & 0.0 & 0.0 & 0.0 & 0.0 & read&0.0 & \textbf{1.0} & 0.0 & 0.0 & 0.0 & 0.0 \\
			0.0 & 0.0 & \textbf{1.0} & 0.0 & 0.0 & 0.0 &start&0.0 & 0.0 & \textbf{1.0} & 0.0 & 0.0 & 0.0 \\
			0.0 & 0.0 & 0.0 & \textbf{1.0} & 0.0 & 0.0 &stop&0.0 & 0.0 & 0.0 & \textbf{1.0} & 0.0 & 0.0 \\
			0.0 & 0.0 & 0.0 & 0.0 & \textbf{1.0} & 0.0 &image&0.0 & 0.0 & 0.0 & 0.0 & \textbf{1.0} & 0.0 \\
			0.01 & 0.0 & 0.0 & 0.0 & 0.0 & \textbf{0.99} &text&0.0 & 0.0 & 0.0 & 0.0 & 0.0 & \textbf{1.0} \\
			\bottomrule
		\end{tabular}
	\end{center}
	\begin{center}
		\begin{tabular}{ccccccccccccc}
			\multicolumn{6}{c}{iOS (From)} & & \multicolumn{6}{c}{iOS (To)}\\
			%\cline{1-7} \cline{9-15}
			\cline{1-6} \cline{8-13}
			control & read & start & stop & image & text &  & control & read & start & stop & image & text \\
			\midrule
			\textbf{1.0} & 0.0 & 0.0 & 0.0 & 0.0 & 0.0 &control &\textbf{0.98} & 0.0 & 0.0 & 0.0 & 0.0 & 0.02 \\
			0.0 & \textbf{1.0} & 0.0 & 0.0 & 0.0 & 0.0 &read&0.0 &\textbf{0.0} & 1.0 & 0.0 & 0.0 & 0.0 \\
			0.0 & 0.0 & \textbf{1.0} & 0.0 & 0.0 & 0.0 &start&0.0 & 0.0 & \textbf{1.0} & 0.0 & 0.05 & 0.0 \\
			0.0 & 0.0 & 0.0 & \textbf{1.0} & 0.0 & 0.0 &stop&0.01 & 0.0 & 0.0 & \textbf{0.99} & 0.0 & 0.0 \\
			0.0 & 0.0 & 0.01 & 0.0 & \textbf{0.99} & 0.0 &image&0.01 & 0.0 & 0.0 & 0.0 & \textbf{0.99} & 0.0 \\
			0.0 & 0.0 & 0.0 & 0.0 & 0.0 & \textbf{1.0} &text&0.01 & 0.0 & 0.0 & 0.0 & 0.04 & \textbf{0.99} \\
			\bottomrule
		\end{tabular}
	\end{center}
	\caption{Confusion matrix for message type classification using iOS and OSX data.}
	\label{tbl:type_confusion}
\end{table*}

In an effort to focus our evaluation, we assume that the OS has already been accurately classified 
such that we have four separate message-type classifiers, one for each combination of OS and direction.  
Each of the classifiers is evaluated using 10-fold cross validation with instances drawn from the respective subsets of the dataset, 
for a total of 1,250 instances per classifier.  Confusion matrices showing the results for OSX and iOS 
are presented in Table  \ref{tbl:type_confusion}.  The accuracy is surprisingly good for both iOS and OSX
given such a simple classification strategy.  As it turns out, all message types can be classified with accuracy exceeding 99\%, except 
for iOS read messages that are easily confused with start messages, as was suggested by \figref{fig:line_types}.

\subsection{Message Attributes}
\label{sec:lang}
\begin{figure*}[t!]
	\footnotesize
	\centering
	\includegraphics[scale=0.45]{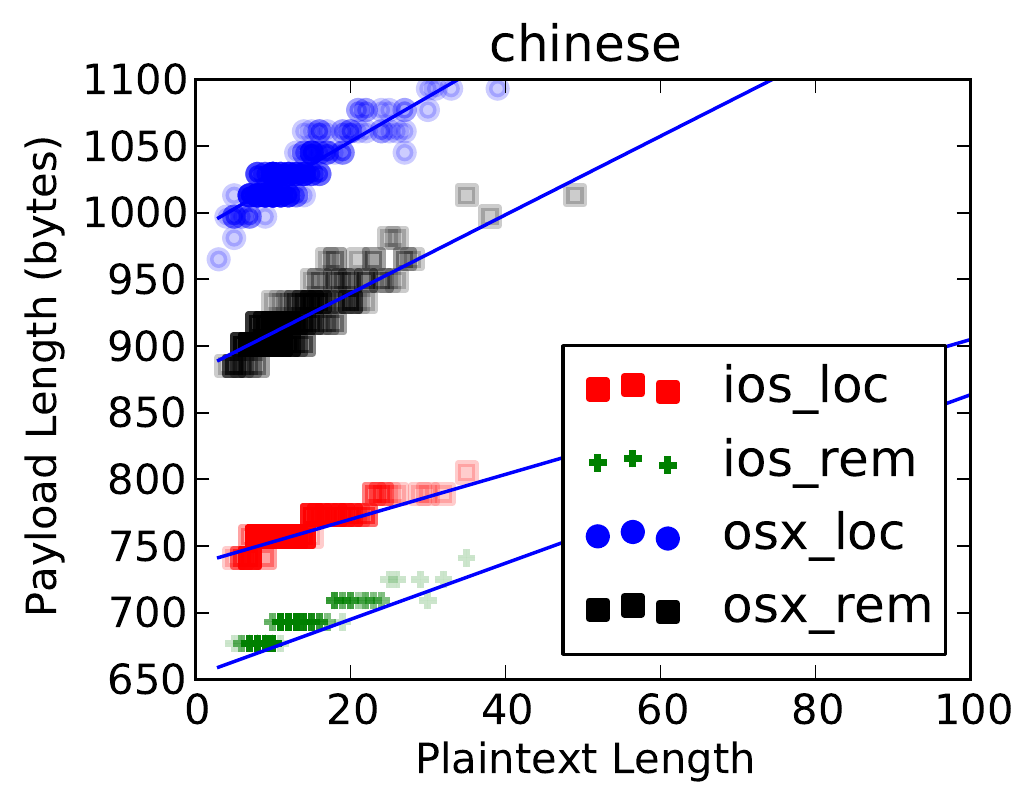}
	\includegraphics[scale=0.45]{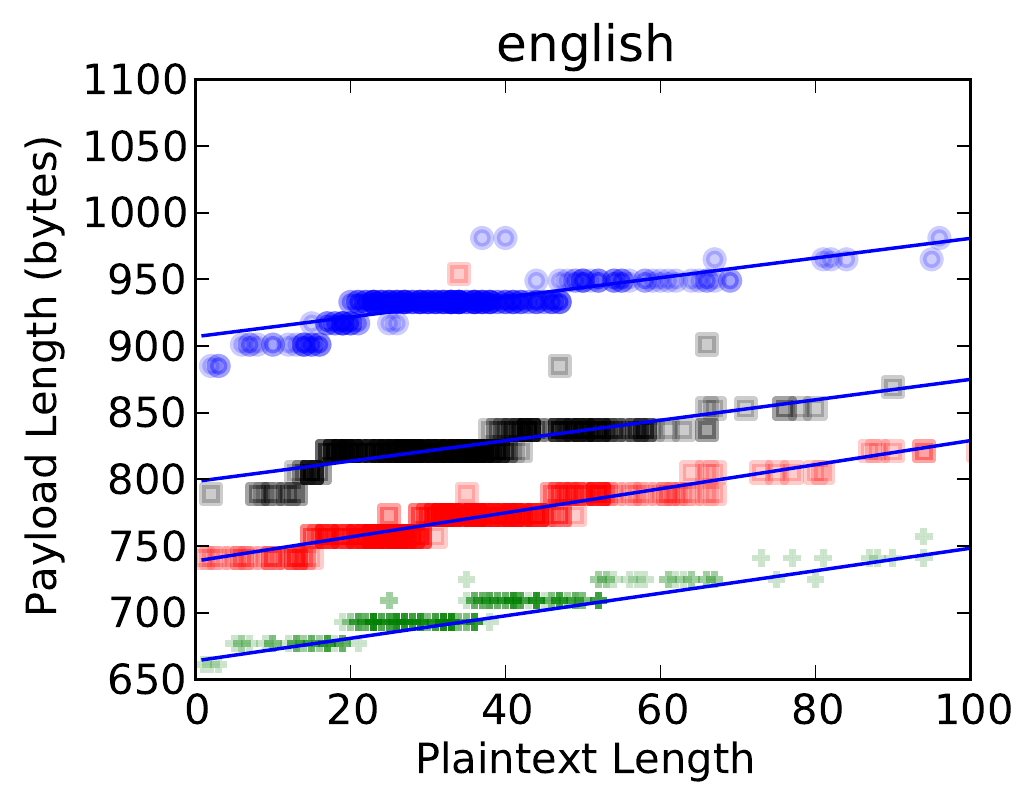}
	\includegraphics[scale=0.45]{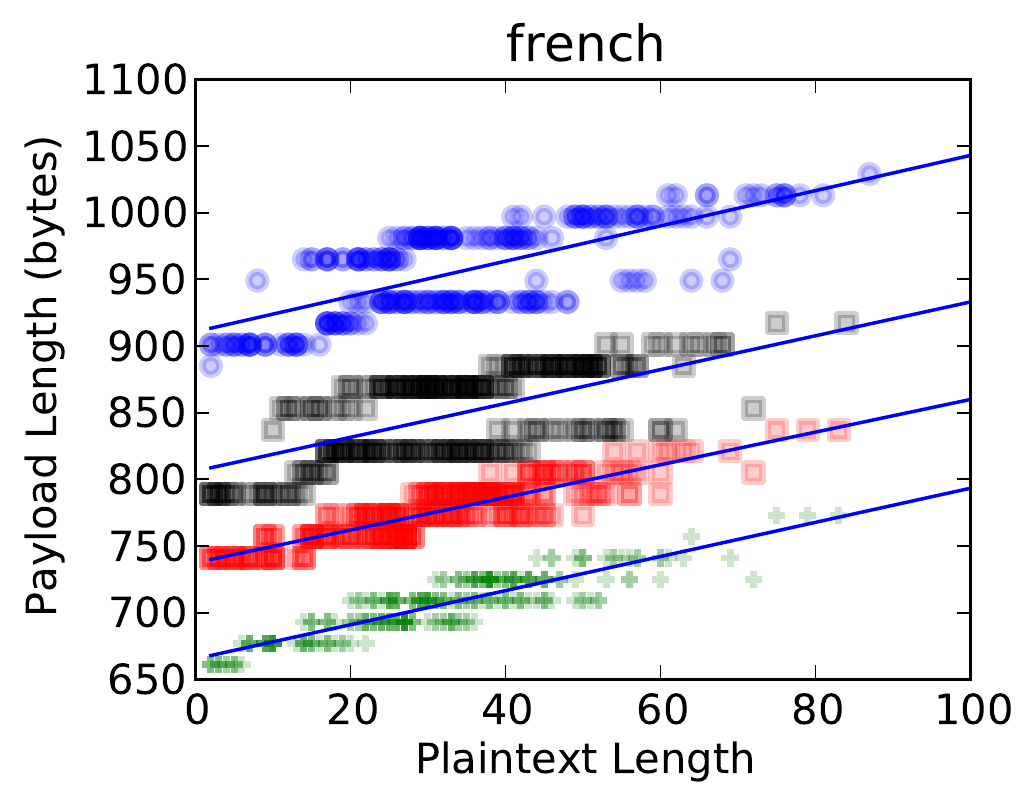}
	\caption{Scatter plots of plaintext message lengths versus payload lengths for three languages in our dataset.}
	\label{fig:language_regression}
\end{figure*}

The final experiment in our analysis of information leakage examines if it is possible to learn more detailed information 
about the contents of messages, such as their language or plaintext length.  The foundation for 
this experiment is built upon the observation that \figref{fig:scatter_os} (in Section \ref{sec:fingerprint}) shows several 
distinct clusters when comparing plaintext message length to payload length.  While the clusters are most prevalent 
in the OSX data, the iOS data also has a similar set of clusters (albeit more compressed).  When we 
separate this data into its constituent languages, as in \figref{fig:language_regression}, the reason for these 
clusters becomes clear.  Essentially, each cluster represents a unique character set used in the language (\eg ASCII, Unicode).  
For languages that use only a single character set, like English (ASCII), Russian (Unicode), or Chinese (Unicode), there is only 
one cluster approximating a linear relationship between plaintext and payload lengths, with a ``stair step" effect at AES block  
boundaries.  The other three languages all use some mix of ASCII and Unicode characters, resulting in an ASCII cluster with 
better plaintext/payload length ratios, and Unicode cluster that requires more payload bytes to encode the plaintext message.  
These graphs also help to answer our question about the possibility of guessing the message lengths, which is supported by the 
approximately linear relationship that appears.  

\begin{figure}
	\footnotesize
	\center
	\begin{subfigure}{0.49\columnwidth}
		\centering
		\includegraphics[scale=0.42]{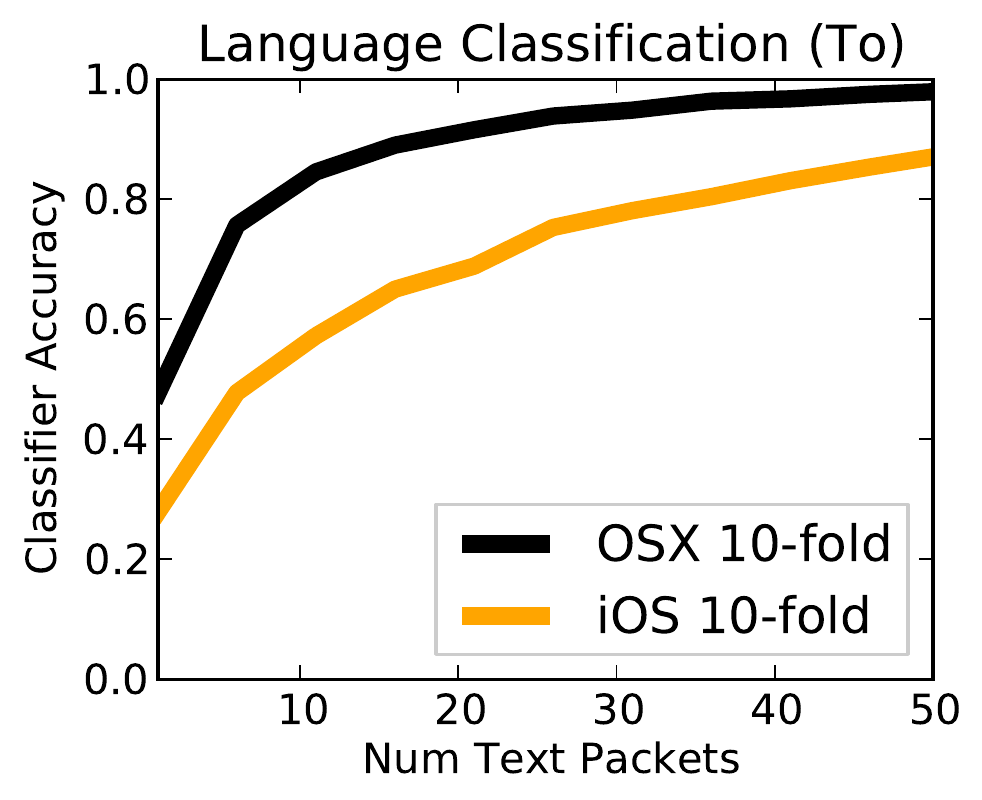}
		\label{fig:os_local}
	\end{subfigure}
	\begin{subfigure}{0.49\columnwidth}
		\centering
		\includegraphics[scale=0.42]{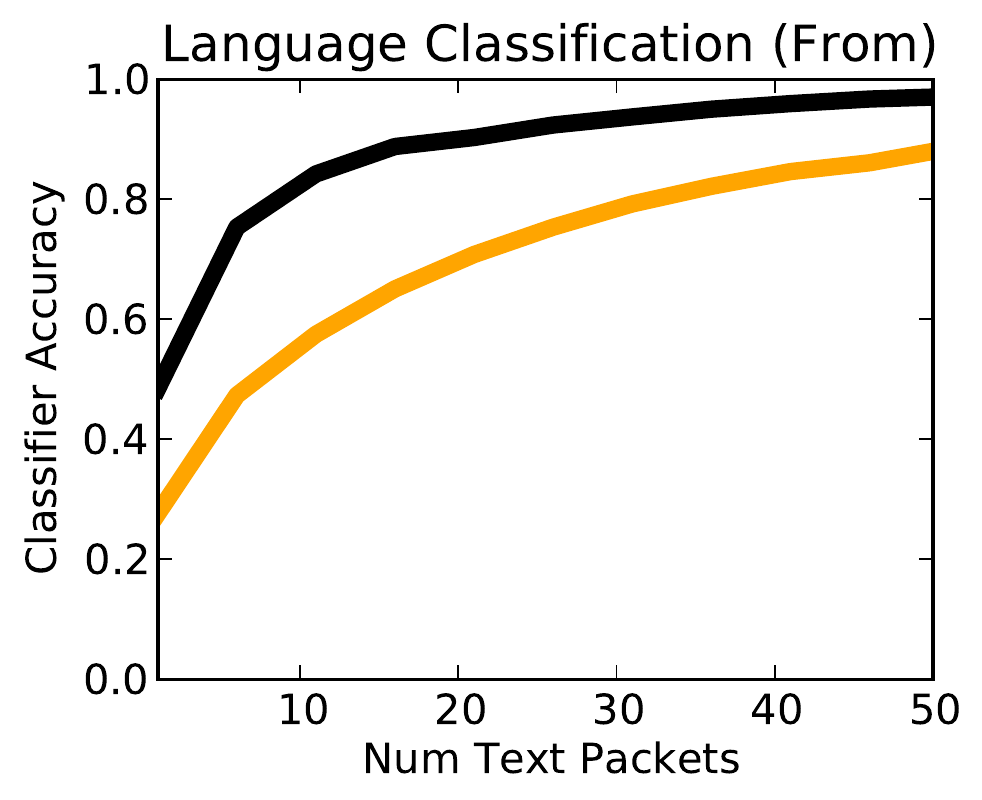}
		\label{fig:os_remote}
	\end{subfigure}
	\caption{Language classification accuracy.}
	\label{fig:lang_results}
\end{figure}

To test our ability to classify these languages, we use the Weka multinomial na\"ive Bayes classifier, with raw counts of each 
length, (packet) direction pair observed so that we can take full advantage of the subtle differences in the distribution.  As with 
previous experiments, we assume that earlier classification stages for OS and message type were 100\% accurate in order to focus 
specifically on this area of leakage.  The results from 10-fold cross validation on 1,024 instances generated from $N=1,2,\ldots,50$ text 
message packets are shown in \figref{fig:lang_results}.  Classification of languages in OSX data is noticeably better than iOS, as we might 
have expected due to compression.  On the OSX data, we achieve an accuracy of over 95\% after 50 packets are observed.  When applied to the iOS data, 
on the other hand, accuracy barely surpasses 80\% at the same number of packets.  However, as the confusion matrices in Table 
\ref{tbl:lang_confusion} show, by the time we sample 100 packets all languages are achieving 
classification accuracies of at least 92\% regardless of the dataset.

Given that language classification can be achieved with high accuracy after a reasonable number of observations, we now move on to determining how 
well we can predict message lengths within those languages.  For this task, we apply a simple linear regression model 
using the payload length as the explanatory variable and the message length as the dependent variable.  The models are 
fitted to the training data using least squares estimation.  Again, we performed 10-fold cross validation with 1,024 instances and calculated the resultant 
absolute error.  In general, the values are small -- an error of between 2 and 11 characters -- when we consider 
that the sentences in the language dataset range from two characters to several hundred, with an average error of 6.27 characters.  
Those languages with multiple clusters, like French and Spanish, 
fared the worst since the linear regression model could not handle the bimodal behavior of the distribution for the multiple character sets.  For completeness, 
we also applied a regression model to the image transfers to and from the Microsoft Azure cloud storage system.  The regression model 
was extremely accurate for the attachments, with an absolute error of less than 10 bytes.

\ignore{
\begin{figure*}[t!]
	\footnotesize
	\center
	\includegraphics[scale=0.5]{images/scatter_linear[ios_loc]}
	\includegraphics[scale=0.5]{images/scatter_linear[ios_rem]}
	\includegraphics[scale=0.5]{images/scatter_linear[osx_loc]}
	\includegraphics[scale=0.5]{images/scatter_linear[osx_rem]}
	\caption{Distribution of payload lengths by language.}
	\label{fig:line_lang}
\end{figure*}
}

\begin{table*}
	\scriptsize
	\begin{center}
		\begin{tabular}{ccccccccccccc}
			\multicolumn{6}{c}{OSX (From)} & & \multicolumn{6}{c}{OSX (To)}\\
			\cline{1-6} \cline{8-13}
			chinese & english & french & german & russian & spanish&  & chinese & english & french & german & russian & spanish\\
			\midrule
			\textbf{1.0} & 0.0 & 0.0 & 0.0 & 0.0 & 0.0&chinese &\textbf{1.0} & 0.0 & 0.0 & 0.0 & 0.0 & 0.0\\
			0.0 & \textbf{1.0} & 0.0 & 0.0 & 0.0 & 0.0 &english&0.0 & \textbf{1.0} & 0.0 & 0.0 & 0.0 & 0.0\\
			0.0 & 0.0 & \textbf{0.98} & 0.0 & 0.0 & 0.02 &french&0.0 & 0.0 & \textbf{0.99} & 0.0 & 0.0 & 0.01\\
			0.0 & 0.0 & 0.0 & \textbf{1.0} & 0.0 & 0.0 & german&0.0 & 0.0 & 0.0 & \textbf{1.0} & 0.0 & 0.0\\
			0.0 & 0.0 & 0.0 & 0.0 & \textbf{1.0} & 0.0 & russian&0.0 & 0.0 & 0.0 & 0.0 & \textbf{1.0} & 0.0\\
			0.0 & 0.0 & 0.02 & 0.0 & 0.0 & \textbf{0.98} & spanish&0.0 & 0.0 & 0.0 & 0.0 & 0.0 & \textbf{1.0}\\
			\bottomrule
		\end{tabular}
	\end{center}
	\begin{center}
		\begin{tabular}{ccccccccccccc}
			\multicolumn{6}{c}{iOS (From)} & & \multicolumn{6}{c}{iOS (To)}\\
			\cline{1-6} \cline{8-13}
			chinese & english & french & german & russian & spanish&  & chinese & english & french & german & russian & spanish\\
			\midrule
			\textbf{1.0} & 0.0 & 0.0 & 0.0 & 0.0 & 0.0&chinese &\textbf{1.0} & 0.0 & 0.0 & 0.0 & 0.0 & 0.0\\
			0.0 & \textbf{0.99} & 0.0 & 0.0 & 0.01 & 0.0 &english&0.0 & \textbf{1.0} & 0.0 & 0.0 & 0.0 & 0.0\\
			0.0 & 0.0 & \textbf{0.98} & 0.01 & 0.01 & 0.0 &french&0.0 & 0.0 & \textbf{0.92} & 0.06 & 0.02 & 0.0\\
			0.0 & 0.0 & 0.02 & \textbf{0.97} & 0.01 & 0.0 & german&0.0 & 0.0 & 0.04 & \textbf{0.96} & 0.01 & 0.0\\
			0.0 & 0.01 & 0.01 & 0.0 & \textbf{0.95} & 0.03 & russian&0.0 & 0.0 & 0.02 & 0.0 & \textbf{0.95} & 0.03\\
			0.0 & 0.0 & 0.01 & 0.0 & 0.06 & \textbf{0.94} & spanish&0.0 & 0.0 & 0.01 & 0.0 & 0.07 & \textbf{0.92}\\
			\bottomrule
		\end{tabular}
	\end{center}
	\caption{Confusion matrix for language classification using iOS and OSX data after observing 100 packets.}
	\label{tbl:lang_confusion}
\end{table*}

\ignore{
\begin{table}
	\footnotesize
	\center
	\begin{center}
	\begin{tabular}{lcc}
		\toprule
		&OSX  & iOS \\
		\midrule
		chinese& 	2.35&	4.05 \\
		english&	6.55&	9.22 \\
		french&	10.14&	5.60\\
		german&	8.09&	5.79 \\
		russian&	3.95&	2.97\\
		spanish&	10.35&	6.25\\
		\bottomrule
	\end{tabular}
	\end{center}
	\caption{Avg. absolute error of message length predictions.}
	\label{tbl:lang_rmse}
\end{table}
}

\section{Beyond iMessage}
\label{sec:conclusion}

\begin{figure*}
	\footnotesize
	\centering
	\includegraphics[scale=0.41]{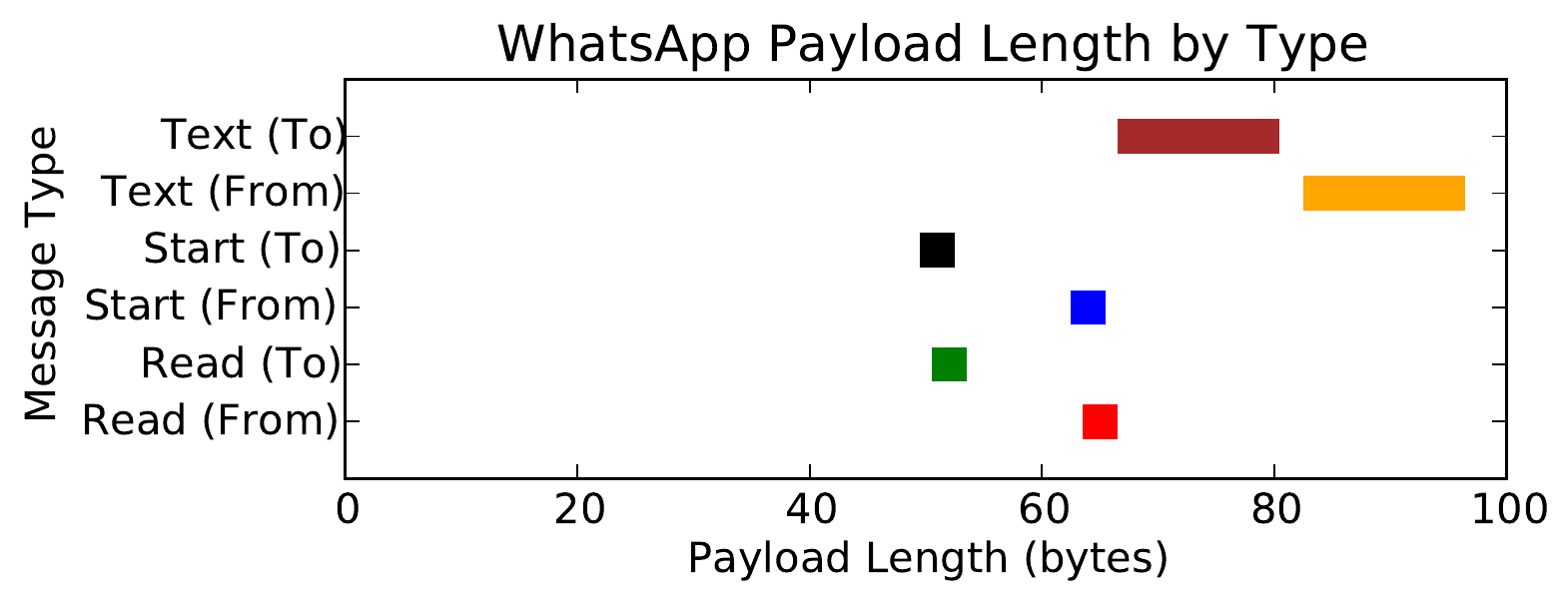}
	\includegraphics[scale=0.41]{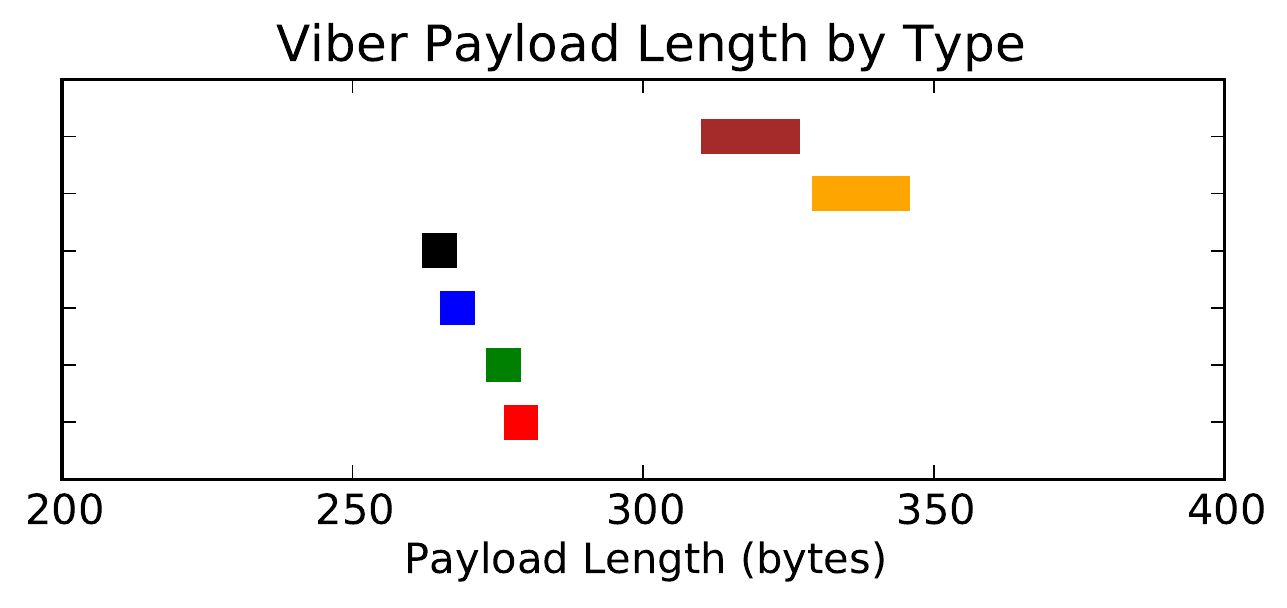}
	\includegraphics[scale=0.41]{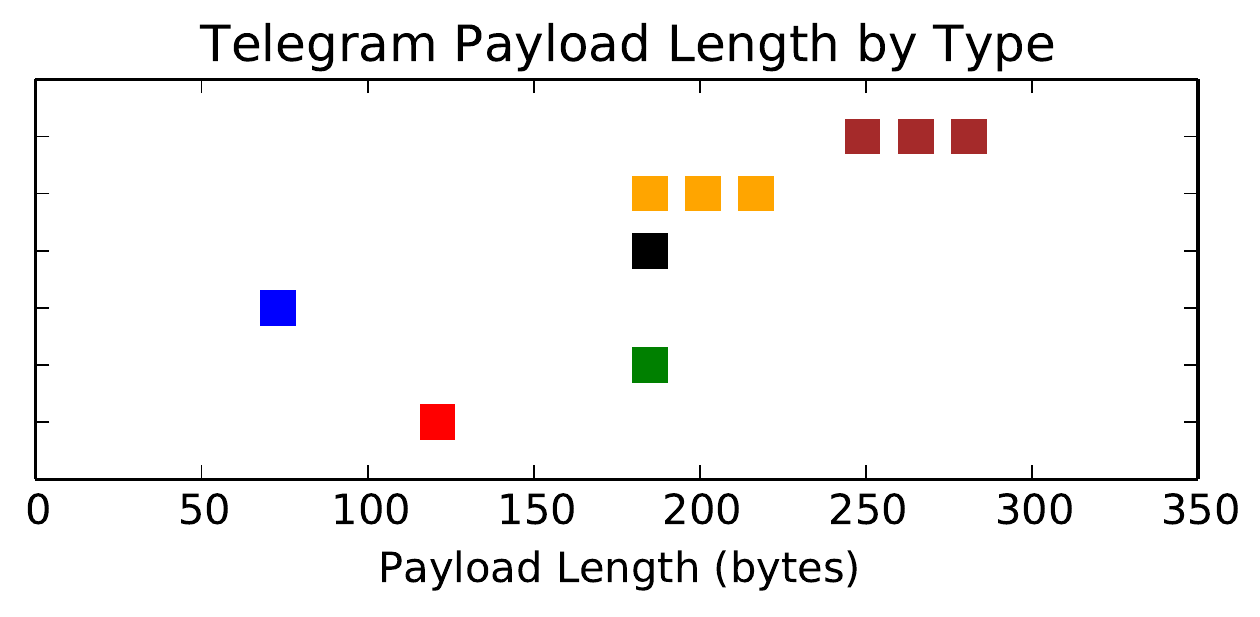}
	\caption{Distribution of payload lengths by type for WhatsApp, Viber, and Telegram.}
	\label{fig:viber_whatsapp_type}
\end{figure*}

\begin{figure*}
	\footnotesize
	\centering
	\includegraphics[scale=0.41]{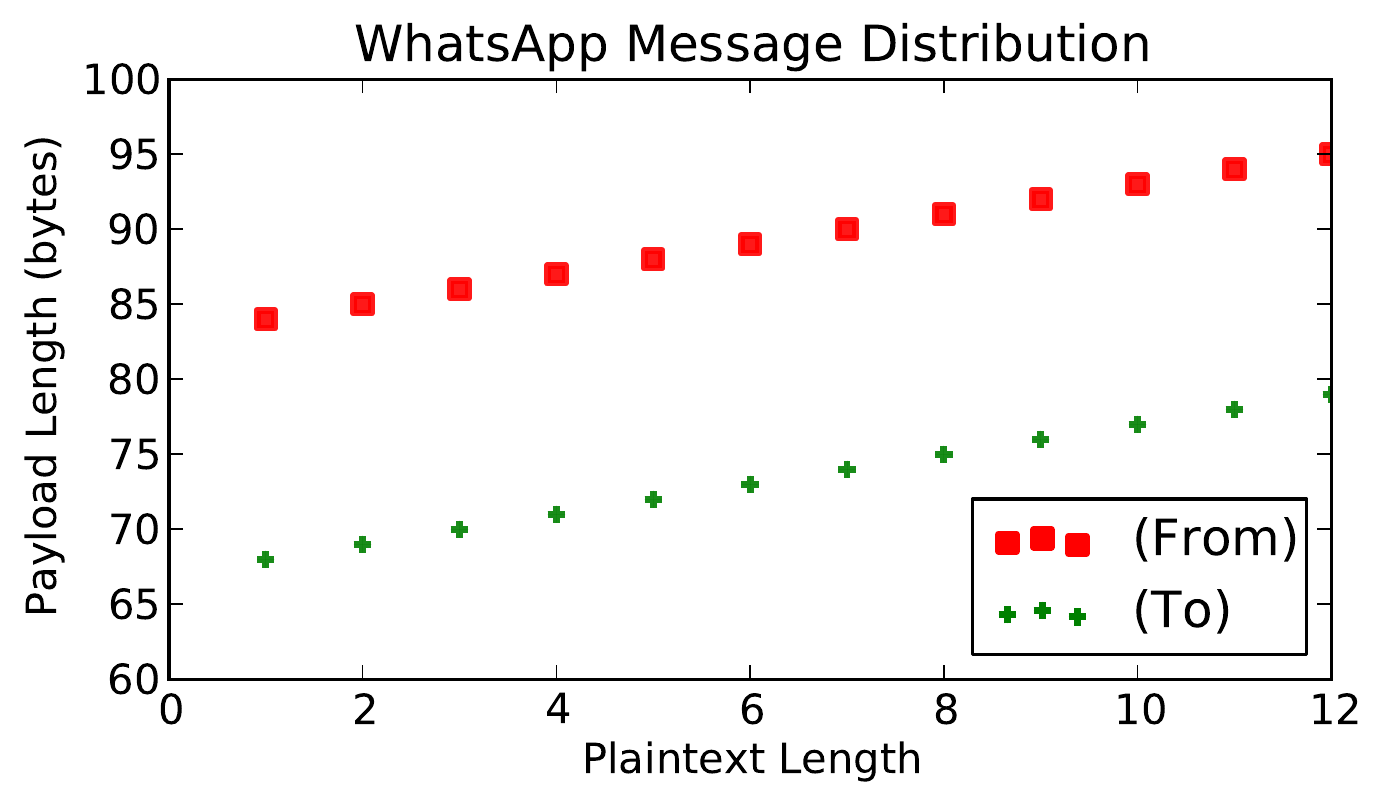}
	\includegraphics[scale=0.41]{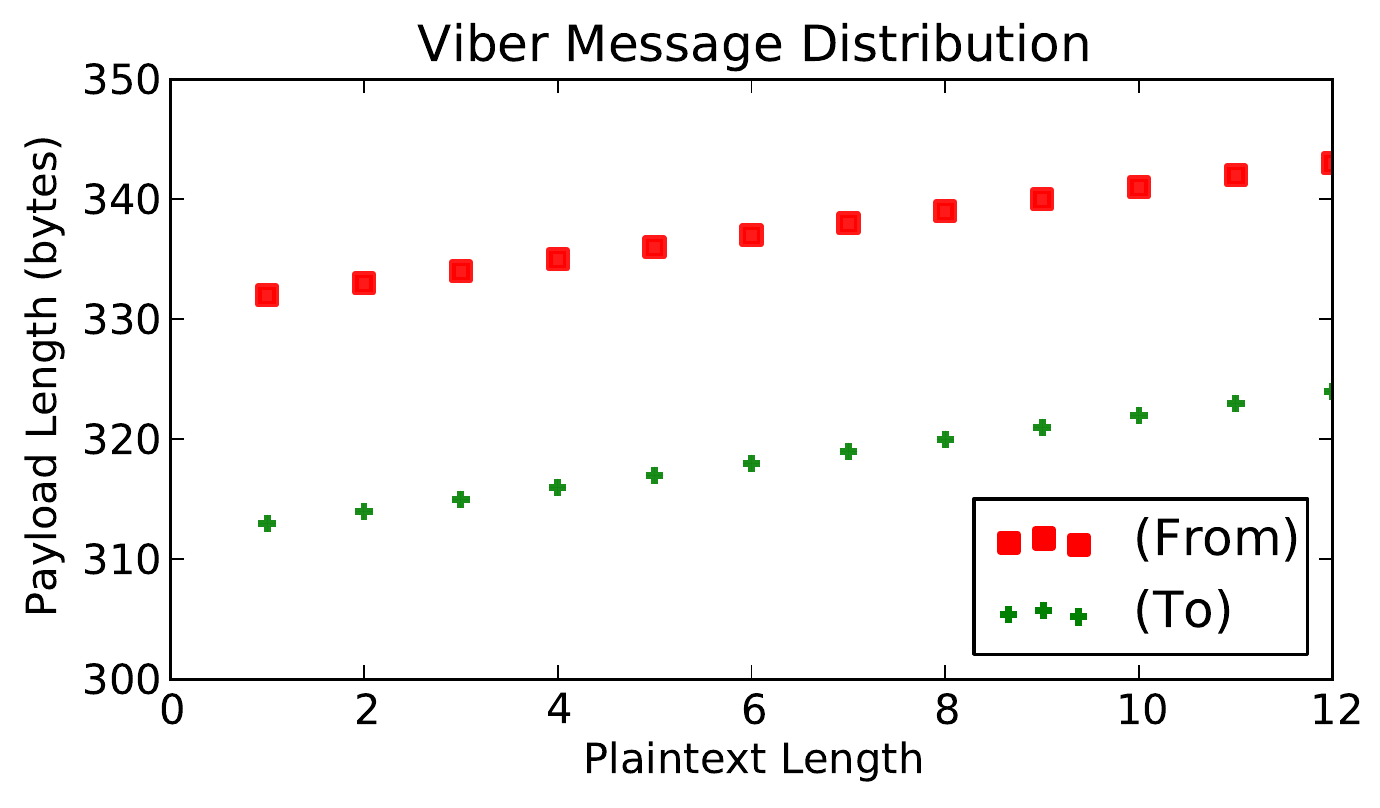}
	\includegraphics[scale=0.41]{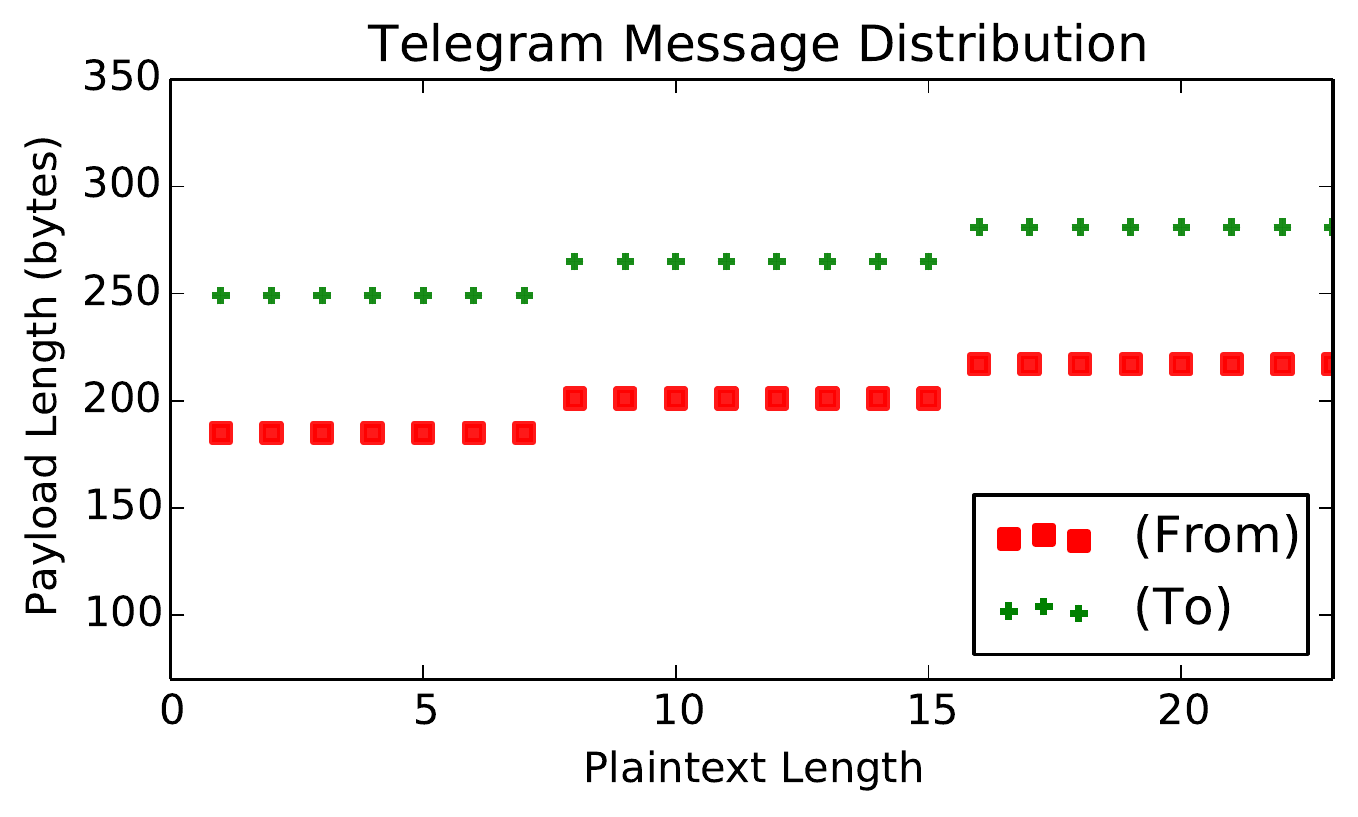}
	\caption{Scatterplot of plaintext message lengths versus payload lengths for WhatsApp, Viber, and Telegram.}
	\label{fig:viber_whatsapp_dist}
\end{figure*}

Thus far, we have focused our attacks exclusively on Apple iMessage, 
however we note that they rely only on the user's interaction with the messaging 
service and a deterministic relationship between those actions and packet sizes.  In effect, the attacks target fundamental 
operations that are common to all messaging services.  To illustrate this concept, we used the same data generation 
procedures described in Section \ref{sec:data_method} to examine the leakage of user actions and message information 
in the WhatsApp, Viber, and Telegram messaging services.  Figure \ref{fig:viber_whatsapp_type} shows the distribution of packet lengths 
associated with the user actions that we have considered throughout this paper for those services.  Just as with Apple iMessage 
(\cf Figure \ref{fig:line_types}), these three messaging services clearly allow us to differentiate fine-grained activities by examining 
individual packet sizes.  Moreover, when we examine the relationship between plaintext message lengths and ciphertext length, as 
in Figure \ref{fig:viber_whatsapp_dist}, there is a clear linear relationship between the two.

Figures \ref{fig:viber_whatsapp_type} and \ref{fig:viber_whatsapp_dist} illustrate two very important concepts in our study.  
First, it shows that the same general strategies used to infer 
user actions, languages, and message lengths can be used across many of the most popular messaging services regardless of their 
individual choices in data encoding, protocols, and encryption.  Second, it is clear that WhatsApp and Viber provide even weaker protection 
against information leakage than iMessage, since there are exact one-to-one relationships between packet sizes and plaintext message lengths.  
Specifically, in Section \ref{sec:types}, we mentioned that Apple iMessage data showed a ``stair step" pattern due to the AES block sizes 
used, which naturally quantizes the output space and adds uncertainty to message length predictions, while Viber and WhatsApp allow us to 
\emph{exactly} predict message length.  Telegram, with its use of end-to-end encryption technology, appears to be very similar to iMessage 
in terms of its payload length distributions.  Therefore, we can expect the accuracy of the attacks will be at least as good as what was demonstrated 
on Apple iMessage traffic.

To mitigate against such privacy failures, 
it is possible to apply standard padding-based countermeasures.  Apple iMessage and Telegram already implement a weak form of countermeasure through 
packet sizes quantized at AES block boundaries.  A much more effective approach, however, would be to add random padding independently to each 
packet up to the maximum observed packet length for each service, thereby destroying any relationship to user actions.  
When implemented on our Apple iMessage data, the random padding methodology reduced all 
of our attacks to an accuracy of 0\% at the cost of 613 bytes (328\%) of overhead per message for iOS and 596 bytes (302\%) for OSX.  
Although the absolute increase in size is rather small, we must consider that services like iMessage handle upwards of 2 billion messages 
every day, which translates to an additional terabyte of network traffic daily.  For the more popular WhatsApp service, a similar increase would incur 
at least 4 terabytes of overhead.  Other countermeasure methods, such as traffic morphing \cite{Wright:Morphing}, may actually provide a more palatable trade-off between 
overhead and privacy.

Overall, the attacks that we have demonstrated raise a number of very important questions about the level of privacy that users can 
expect from these services.  While the exact plaintext content cannot (yet) be revealed, rich metadata can be learned about a user and their 
social network.  In the wake of recent reports of widespread metadata gathering by government agencies \cite{Ackerman:Optic_Nerve, Cohn:Metadata} 
and given the unusually broad impact 
of these attacks on an international user base, it seems reasonable to assume that these types of attacks are a realistic threat that should be 
taken seriously by messaging services.

\small
\bibliography{imessage}

\end{document}